\documentclass{aastex7}
\usepackage{pifont}
\usepackage{amsmath}
\usepackage{amssymb}
\usepackage{graphicx}

\begin{document}

\title{Discussion on the vanishment of solar atmospheric structures during magnetic reconnection}

\author[0000-0002-1286-6931]{Jun Zhang}
\affiliation{School of Physics, Anhui University, Hefei 230601, People's Republic of China; \textcolor{blue}{zjun@ahu.edu.cn}}
\email{zjun@ahu.edu.cn}

\author[0000-0002-5383-1129]{Tao Ding}
\affiliation{School of Materials Science and Engineering, Anhui University, Hefei 230601, People's Republic of China; \textcolor{blue}{dtao@ahu.edu.cn}}
\email{dtao@ahu.edu.cn}

\author[0000-0001-9863-5917]{Yulei Wang}
\affiliation{School of Astronomy and Space Science, Nanjing University, Nanjing 210023, People's Republic of China}
\affiliation{Key Laboratory for Modern Astronomy and Astrophysics (Nanjing University), Ministry of Education, Nanjing 210023, People's Republic of China}
\email{wyulei@nju.edu.cn}

\begin{abstract}

In solar atmosphere, magnetic reconnection alters the topological connectivity, and magnetic energy is released. However, the length change of the reconnecting structures has rarely been reported. To identify the evolution of the topological structures, we search for reconnection events which should satisfy 3 criteria. (1) Each event displays an explicit X-type configuration, and the configuration consists of two sets of independent atmospheric structures, (2) the reconnection process is clearly observed, and (3) the topological connectivity of the structures can be tracked from at least 5 minutes prior to the occurrence of magnetic reconnection to 5 minutes after the reconnection. In this work, 3 events are selected and studied. During the reconnection moment, the total length of the two topological structures in each event shortens suddenly, and the decrements for events 1--3 are 47 Mm, 3.7 Mm, and 8.2 Mm, respectively, implying that partial structures vanish observationally during magnetic reconnection process. Several possibilities about the vanishment, e.g. the shrinkage of the reconnecting structures due to magnetic tension, the bizarre change in the third dimension, and magnetic field annihilation, have been discussed.


\end{abstract}

\keywords{Magnetic reconnection – Sun: activity – Sun: magnetic fields}

\section{Introduction}

It is believed that magnetic reconnection plays an important role in energy release and conversion among magnetized plasma systems \citep{2006ApJ...648L..51S,2006PhPl...13e2119Y,2008ApJ...688..555G}. In solar atmosphere, the distinct change of the topological connectivity of magnetic fields takes place during the reconnection process \citep{2000mare.book.....P,2010RvMP...82..603Y}, and the stored magnetic field energy is rapidly released and converted into the kinetic and thermal energy of plasmas, resulting in a new equilibrium configuration of lower energy. It is reported that various activities, such as flares \citep{2013NatPh...9..489S,2023ApJ...947...67R}, jets \citep{1992PASJ...44L.173S,2023A&A...670A..56B}, and coronal mass ejections \citep{2018SSRv..214...46G,2023A&A...672A..23S}, are associated with the reconnection.

The traditional magnetic reconnection scenario is described as following. When two opposite magnetic field lines close enough, the particles along with the two lines cannot distinguish one line from the other one, so the electron on a line can be captured by the other line, or even loses completely. Under this condition, the magnetic line has lost its own identity, which means that we cannot identify the magnetic line. The area where the magnetic force lines lose their identity is called the diffusion region \citep{1984GMS....30..379D}. A current sheet forms, and magnetic energy converts into heat and kinetic energy by Ohmic dissipation in the diffusion region \citep{1946Natur.158...81G,1958IAUS....6..123S}. The topology of the magnetic field in this region can be changed, e.g. magnetic fields can be broken and rejoined \citep{2010RvMP...82..603Y}. Once this change occurs, the magnetic line crossing over this diffusion zone is no longer the original magnetic line, and the connection is reorganized. The whole process of putting this magnetic line into the diffusion zone, losing identity, reconnecting, and finally crossing over the diffusion zone is called magnetic reconnection. Obviously, magnetic reconnection is accompanied by changes in the magnetic field topology, thus resulting in the rapid release of magnetic field energy \citep{1957JGR....62..509P,1958IAUS....6..123S,1964NASSP..50..425P,2010RvMP...82..603Y}. However, the topological evolution of the reconnecting fields before, during, and after the magnetic reconnection has not been considered seriously.  

In this work, we quantitatively measure for the first time the topological changes of magnetic structures during reconnection from observations, and then put forward a new topological evolution scenario to interpret the observations.

\section{Observations}

To obtain reliably the lengths of the reconnected structures, we focus on the reconnection events which have explicit X-type configurations. The X-type configuration in each event consists of two sets of independent atmospheric structures. Then, the reconnection process is clearly observed, and the topological connectivity of the structures can be tracked from at least 5 minutes prior to the occurrence of magnetic reconnection to 5 minutes after the reconnection. We check all reported literature based on the data from the Atmospheric Imaging Assembly \citep[AIA;][]{2012SoPh..275...17L} onboard the Solar Dynamical Observatory \citep[SDO;][]{2012SoPh..275....3P} and the New Vacuum Solar Telescope \citep[NVST;][]{2014RAA....14..705L}. The numbers of the magnetic reconnection events related to the SDO/AIA and NVST H${\alpha}$ are 48 and 11 \citep{2024ApJ...964...16D, 2024ApJ...964...58D}, respectively. Finally, 3 events (e.g. events 1--3 in Figures \ref{fig1}--\ref{fig3}) are studied in detail. Event 1 is based on the SDO/AIA full-disk multiband extreme ultraviolet (EUV) observations, with the spatial sampling and the time cadence of 0.$''$6 pixel$^{-1}$ and 12 s, respectively. Events 2 and 3 utilize the joint observations from the SDO/AIA and NVST H${\alpha}$. \citet{2024ApJ...964...16D} focused on the evolution of photospheric magnetic fields at the footpoints of 16 magnetic reconnection events. This work studies the variation of reconnected structures, based on the 3 reconnection events, and events 2 and 3 are from the 16 events in \citet{2024ApJ...964...16D}. The NVST H${\alpha}$ raw data (Level 0) are applied to calibrate the Level 1 data, and then the speckle masking method is carried out to derive the Level 1+ data. These H${\alpha}$ data have the spatial resolutions of 0.$''$163 (event 2) and 0.$''$33 (event 3), as well as time cadences of 12 s and 49 s, separately. Finally, the cross-correlation method is employed to coalign the SDO and NVST images.

\section{Results}

\begin{figure}
\plotone{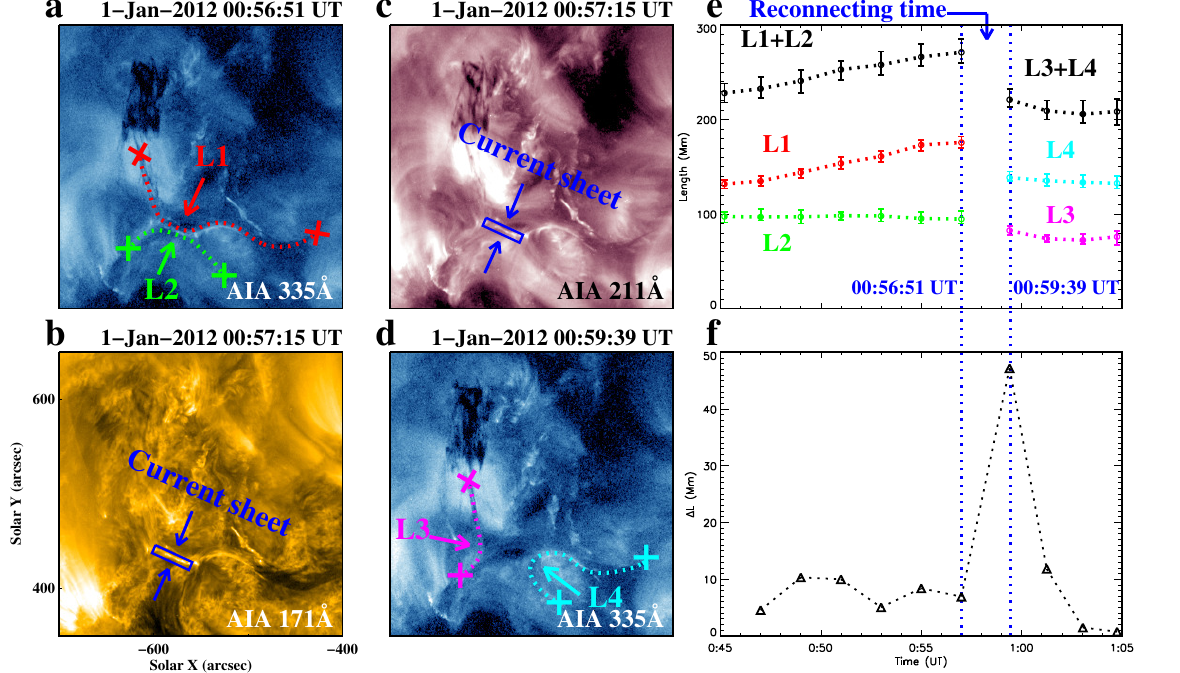}
\centering
\caption{A magnetic reconnection event on 1 January 2012 \citep[Event 1;][]{2016NatPh..12..847L}. Panels (a) and (d), the SDO/AIA 335 Å images. Panels (b) and (c), the current sheet in SDO/AIA 171 (211) Å image. The red (L1) and green (L2) dotted lines in panel (a) display two independent atmospheric structures prior to the occurrence of magnetic reconnection, while the pink (L3) and cyan (L4) ones in panel (d) show the newly-formed structures which consist of the residues of L1 and L2 after the reconnection. The blue rectangles in panels (b) and (c) exhibit the current sheet. Panel (e), the lengths of the topological structures L1, L2, L3 and L4, as well as the length sums of L1 plus L2 and L3 plus L4. Panel (f), the length sum variation of L1 plus L2 prior to the reconnection, and L3 plus L4 after the reconnection. The two blue dotted lines in panels (e) and (f) denote the start (00:56:51 UT) and end (00:59:39 UT) time of the reconnection, respectively. To display the exchange of topological connectivity of magnetic field lines during magnetic reconnection process, an animation of the SDO/AIA 335 Å observations from 2012 January 1 00:46:03 UT to 01:03:15 UT, is available. The duration of this animation is 1 s.
\label{fig1}}
\end{figure}

Figure \ref{fig1} shows a magnetic reconnection event \citep{2016NatPh..12..847L} which happens between a filament L1 and a set of EUV loop L2 (Figure \ref{fig1}(a)). As for each independent atmospheric structure involving in the reconnection event, we label two feature points as the two ends of the structure, and mark them with the “X” symbols (see Figures \ref{fig1}(a) and (d)). The two points must be present throughout the whole reconnection process and remain stationary. By tracking the main body between the two points prior to the reconnection process, we obtain the lengths of L1 and L2. During the reconnection, the connectivity of the two structures (L1 and L2) cannot be determined in the reconnection region, as the signal of a current sheet (Figures \ref{fig1}(b) and (c)) in the region is stronger. 168 s after the reconnection, L3 and L4, which consist of the residues of L1 and L2, are formed (Figure \ref{fig1}(d)), and the lengths of L3 and L4 are also measured. To determine the length error of each structure, we measured the length of each structure for 10 times, and the mean square deviation is considered as the error. We notice that the length of L1 (L2) prior to the reconnection and the length of L3 (L4) after the reconnection all changed gradually (Figure \ref{fig1}(e)). While we focus on the length sum of L1 and L2, as well as the sum of L3 and L4, it appears that there is a sudden change during the reconnection process (Figure \ref{fig1}(f)). The length sum of L3 and L4 is 47 Mm shorter than the length sum of L1 and L2, meaning that partial structure of L1 and L2 is lost observationally during the reconnection. 

\begin{figure}
\plotone{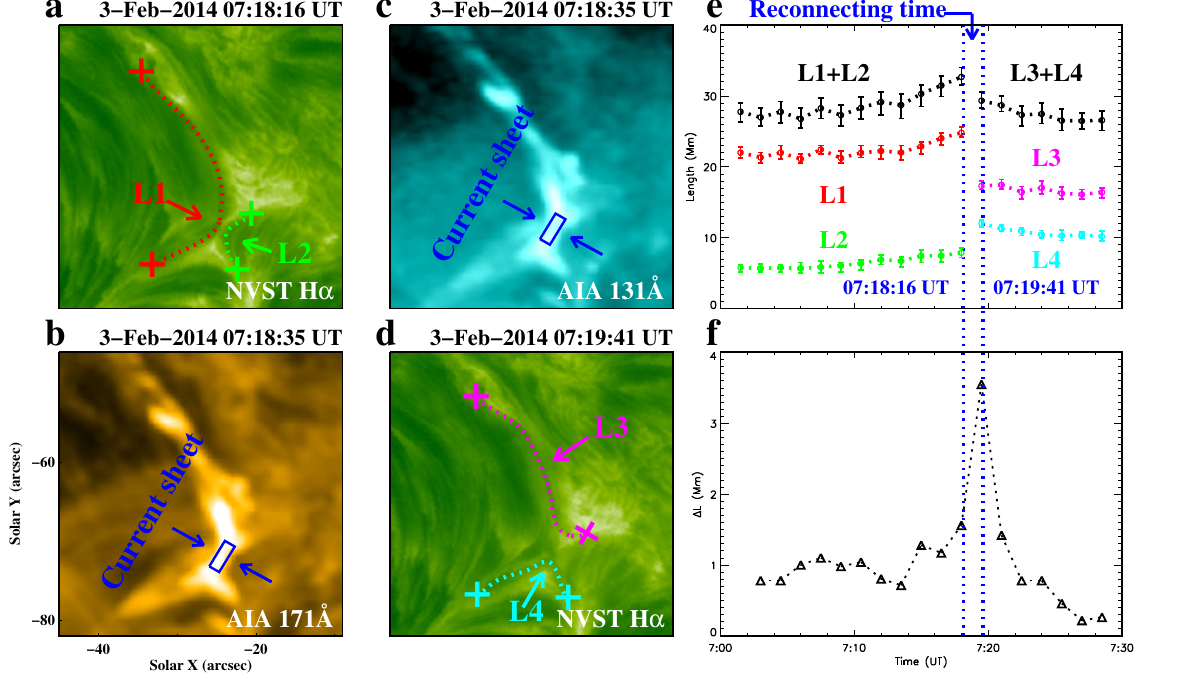}
\centering
\caption{A magnetic reconnection event on 3 February 2014 \citep[Event 2;][]{2015ApJ...798L..11Y}. Panels (a) and (d), the NVST H$\alpha$ images. Panels (b) and (c), The SDO/AIA 171 (131) Å image. The red (L1) and green (L2) dotted lines in panel (a) display two independent atmospheric structures prior to the occurrence of magnetic reconnection, while the pink (L3) and cyan (L4) ones in panel (d) show the newly-formed structures which consist of the residues of L1 and L2 after the reconnection. The blue rectangles in panels (b) and (c) exhibit the current sheet. Panel (e), The lengths of the topological structures L1, L2, L3, and L4, as well as the length sums of L1 plus L2 and L3 plus L4. Panel (f), the length sum variation of L1 plus L2 prior to the reconnection, and L3 plus L4 after the reconnection. The two blue dotted lines in panels (e) and (f) denote the start (07:18:16 UT) and end (07:19:41 UT) time of the reconnection, respectively. To display the exchange of topological connectivity of magnetic field lines during magnetic reconnection process, an animation of the NVST H$\alpha$ observations from 2014 February 3 07:06:10 UT to 07:19:05 UT, is available. The duration of this animation is 1 s.
\label{fig2}}
\end{figure}

\begin{figure}
\plotone{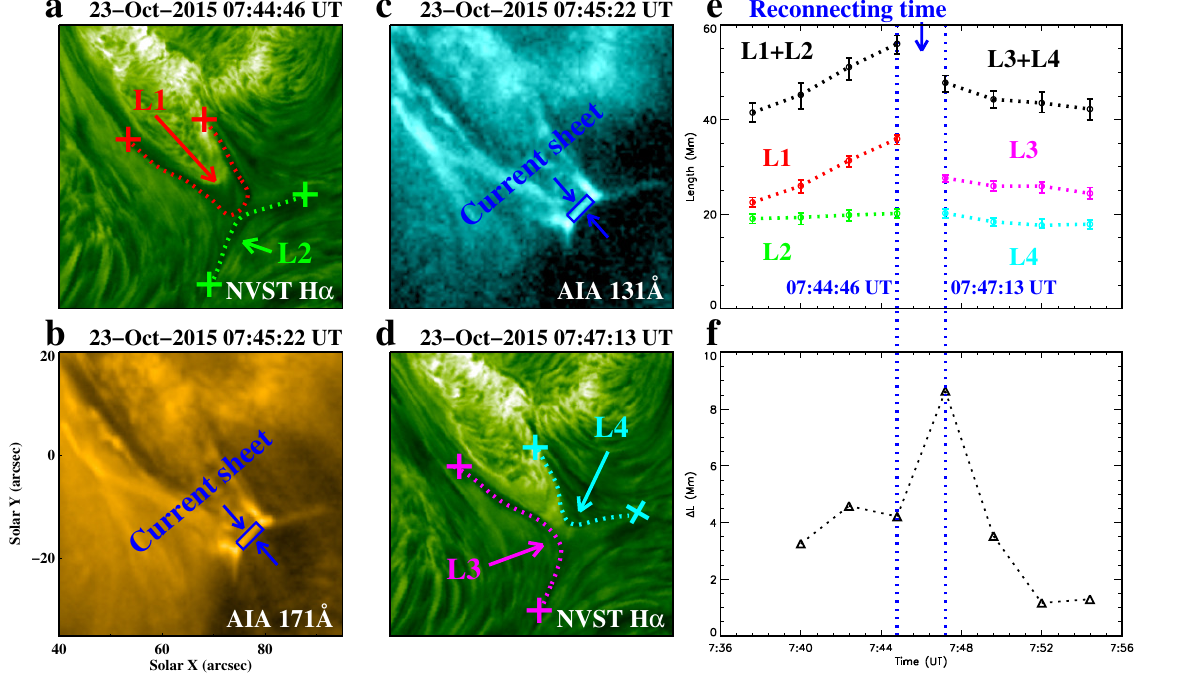}
\centering
\caption{Similar to Figure \ref{fig2}, but for event 3 on 23 October 2015 \citep{xue2020}. The length sum of L3 and L4 is 8.2 Mm shorter than the length sum of L1 and L2. The reduced length is about 18\% of the length sum ($\sim$46 Mm). To display the exchange of topological connectivity of magnetic field lines during magnetic reconnection process, an animation of the NVST H$\alpha$ observations from 2015 October 23 07:37:26 UT to 07:52:07 UT, is available. The duration of this animation is 1 s.
\label{fig3}}
\end{figure}

Figure \ref{fig2} displays the second magnetic reconnection event \citep{2015ApJ...798L..11Y} observed by NVST and SDO AIA on Feb. 3, 2014. This reconnection took place between two chromospheric fibrils L1 and L2 (Figure \ref{fig2}(a)). During the reconnection, a current sheet is observed by AIA 131 and 171 Å wavelengths (Figures \ref{fig2}(b) and (c)). Then, a residue of L1 and a residue of L2 connect to form a new topological structure L3, as well as the other residue of L1 and the other residue of L2 to form L4 (Figure \ref{fig2}(d)). The topological evolution of L1-L4 is similar to the first event in Figure \ref{fig1}. Although the length sum of L3 and L4 is only 3.7 Mm shorter than the length sum of L1 plus L2, the reduced length is still reaches 14\% of the length sum ($\sim$27 Mm). The third event \citep{xue2020} is displayed in Figure \ref{fig3}. Similar to events 1 and 2, the reduced length (8.2 Mm) is about 18\% of the length sum ($\sim$46 Mm).

\section{Discussion}

In this study, we investigate 3 magnetic reconnection events. All the reconnection events display that while magnetic reconnection took place between two set of magnetic fields, a segment of the reconnection structures was destroyed and vanished, and the remainder still existed. The remainder of one set of magnetic field reconnected with the remainder of the other set of field, and a new structure formed. These observations suggest that the topological connectivity is destroyed in reconnection process.

During the magnetic reconnection process, the fields are reconnected in the diffusion region, and magnetic energy is released, thus heating the plasma. On the other hand, we still have no pictures about the evolution of magnetic fields in the diffusion region. Simulations have shown that the reconnection region can be highly turbulent \citep{1999ApJ...517..700L,2009ApJ...700...63K,2016ApJ...818...20H,2022PhPl...29l2902H}, where the fields are chaotic. The turbulent magnetic fields improve the reconnection rate and might also be related to the fine observational phenomena. Besides, another common structure in the diffusion region is magnetic island. It has been observed on the magnetopause, in the magnetotail, and in coronal current sheets. The suggested theory for the island formation is tearing instability \citep{1963PhFl....6..459F}. Both simulations \citep{2006Natur.443..553D} and observations \citep{2008NatPh...4...19C} indicate that the islands are correlated with highly energetic electrons. Electrons are significantly accelerated at the neutral sheet and the subsequent X-line \citep{2009AIPC.1144...15S}. Island contraction in a flare current sheet is a promising candidate for electron acceleration in solar eruptions \citep{2016ApJ...820...60G}. When electrons are trapped inside the islands, they are energized continuously by the reconnection electric field prevalent in the reconnection diffusion region \citep{2010JGRA..115.8223O}. The magnetic island can form secondary (tiny) island due to further tearing instability \citep{2001EP&S...53..473S}. At certain condition, the island can also coalesce with others to generate a larger one as a result of coalescence instability \citep{2008ApJ...689..572B}. These results indicate that the magnetic connectivity is freely exchanged in the diffusion region. \citet{2015ApJ...808..181K} present multi-spacecraft observations of magnetic island merging and particle energization in the absence of other sources, providing support for theory and simulations that show particle energization by reconnection related processes of magnetic island merging and contraction. In the Earth's magnetospherereport, it is reported that there is a connection between energetic electrons and magnetic islands during reconnection \citep{2008NatPh...4...19C,2008GeoRL..3519109P}. Moreover, a consequence of the coalescence instability results in the coalescence of magnetic islands, and energizes electrons \citep{2010JGRA..115.8223O}.

The observed vanishment of the atmospheric structures may result from several possibilities, e.g. the structure evolution, the bizarre change in the third dimension (along the line of sight), and the true vanishment. About the structure evolution, we notice from all the three events that the lengths undergo gentle changes before and after reconnection. Especially after the reconnection, the length sum decreases monotonously. This phenomenon may result from the shrinkage of the reconnecting structures due to the magnetic tension. On the other hand, the suddenly shortening with a big range during reconnection is quite different from the gentle changes. We suggest that the suddenly shortening (vanishment) may have nothing to do with the magnetic tension.

For the bizarre change in the third dimension, we cannot rule out completely this possibility, as the result is obtained from two dimensional plane images. If there is a bizarre change, it can be expected that some information about the change will come out in the subsequent series of the two dimensional images. In fact, no any information is detected, implying  that the bizarre change in the third dimension is not the main reason of the vanishment. 

On this condition, we make bold to put forward a new scenario. The observational vanishment is a true vanishment, i.e. partial structure disappears. The data from the Magnetospheric Multiscale Spacecraft reveal that magnetic field annihilation takes place in Earth's magnetotail \citep{2022JGRA..12730408H}. Magnetic energy converts into electronic energy within a small scale current sheet. Furthermore, a fully kinetic simulation also manifests that small scale magnetic islands can be formed in an elongated electron diffusion region, and fast annihilation occurs in the islands. In a turbulent collisionless plasma region, the annihilation helps with magnetic energy dissipation, and plays a key role in releasing magnetic energy to heat flare plasmas and accelerate particles \citep{2021ApJ...923..227W}. 

It is acceptable that magnetic reconnection releases magnetic energy, and the magnetic energy is closely related to the volume of the magnetic field. We anticipate that the release of magnetic energy reduces the volume of the corresponding magnetic field, so the vanishment of the magnetic structure will correspond to the magnetic energy release. We consider event 1 as an example to estimate the magnetic energy of the vanished structure. The magnetic energy ($E$) can be calculated with $E$=${B{^2}V}/{8\pi}$ \citep{2007ApJ...655L.117S,2012ApJ...748...77S}, where $B$ and $V$ represent the magnetic field strength and the volume of the vanished structure, respectively. During the reconnection, about 8 filament fibrils could be detected (details refer to \citet{2024ApJ...970L...7D}). We assume that each fibril is a cylinder with a radius of $R$ and a vanished length of $L$, then the volume of the vanished structures is $V=8L×\pi×R{^2}$. Here $L$ is the lendth of the vanished structure ($4.7\times10^{9} cm$). Based on the observations, the mean radius $R$ is about $1.6\times10^{8} cm$. For the magnetic field strength, we use a value of $B$=75 G \citep{2016NatPh..12..847L}. The magnetic energy of the vanished structure is $6.8\times10^{29} erg$, which is larger than the current sheet energy, i.e. $1.5\times10^{27} erg s^{-1}\times168$ $s$ = $2.5\times10^{29} erg$ \citep{2016NatPh..12..847L}.

\begin{figure}
\plotone{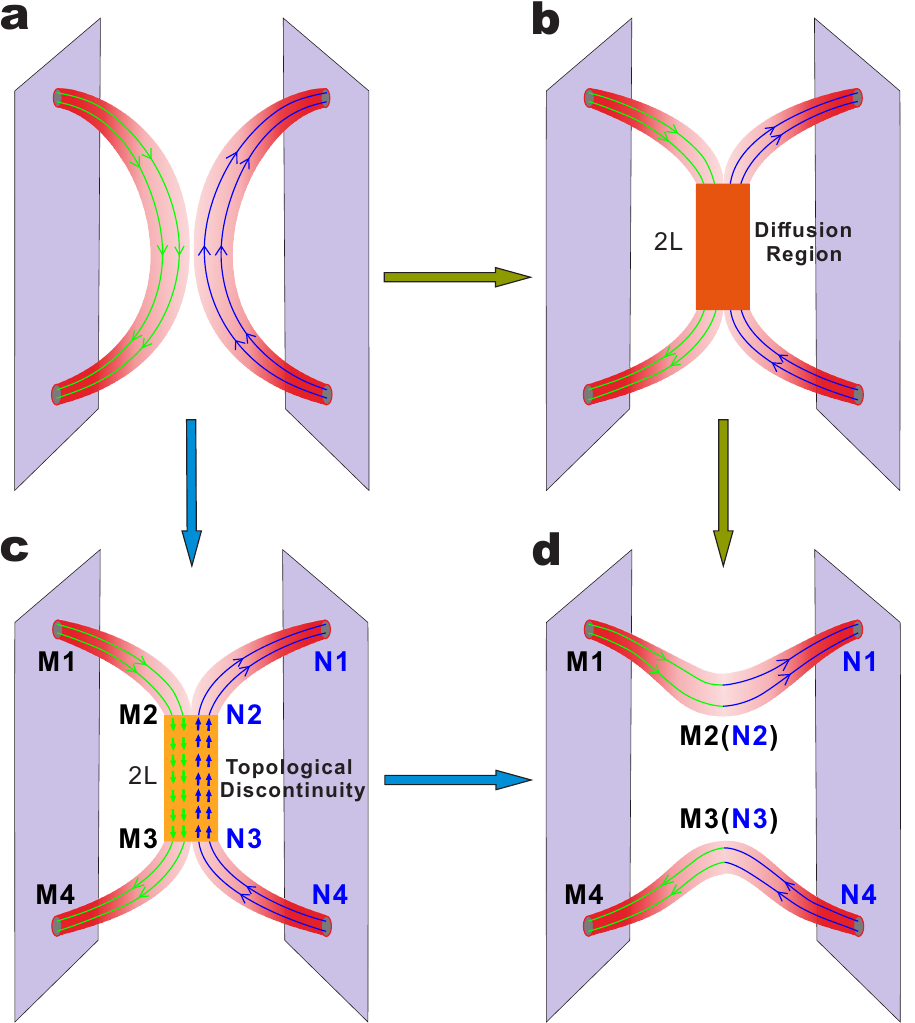}
\centering
\caption{Schematic drawings illustrating the reconnection process. Panels (a), (b), and (d), the traditional reconnection images. In the diffusion region, magnetic energy converts into heat and kinetic energy by Ohmic dissipation. Panel (c), a new idea about the change of the magnetic structures. While two set of magnetic fields approach together in the 2L region (M2-M3 and N2-N3), the magnetic fields fragment (shown by short arrows). The dissipation of these fragmented fields results in the vanishment of the partial structures. The residues of the magnetic fields form two new structures M1-M2(N2)-N1 and M4-M3(N3)-N4, respectively.
\label{fig4}}
\end{figure}

To depict the pictures about the evolution of magnetic fields in the diffusion region, and to interpret the vanishment of partial structures during magnetic reconnection, we exhibit several schematic diagrams in Figure \ref{fig4} to illustrate the reconnection process. Figures \ref{fig4}(a), (b), and (d) display the traditional reconnection images which do not take into account the change of magnetic structures during reconnection. Figure \ref{fig4}(c) shows a new idea about the magnetic structure change. While two sets of magnetic fields (with one set M1-M2-M3-M4, and the other one set N1-N2-N3-N4) approach and close together, e.g. in the 2L region (M2-M3 and N2-N3), the fields display topology discontinuity and fragmentation, then the partial structures M2-M3 and N2-N3 vanish. Outside the 2L region, the residues of the magnetic fields reconnect, and form new structures (e.g. M1-M2(N2)-N1 and M4-M3(N3)-N4 in Figure \ref{fig4}(d)). The relevant conclusions still need to be further confirmed by observations and plasma experiments. In addition, a rigorous mathematical derivation also needs to be perfected.

\begin{acknowledgments}

The observations used in this work are provided by the SDO and NVST teams. This work is supported by the National Natural Science Foundation of China (12573054) and the Anhui Project (Z010118169).

\end{acknowledgments}

\clearpage

\clearpage
\bibliography{reference}{}
\bibliographystyle{aasjournal7}

\end{document}